# The Preponderance of Electric Vehicles and the Availability of "Green" Electricity

Theodore Modis*

* Address reprint requests to:
Theodore Modis
Growth Dynamics
Via Selva 8
6900 Massagno
Lugano, Switzerland.
Tel. 41-91-9212054, E-mail: tmodis@yahoo.com

**Abstract**

The main advantage of electric vehicles, namely, their non-polluting operation, is compromised when the electricity they use comes from burning fossil fuels. The number of electric vehicles on the road is growing much faster than the amount of "green" electricity produced. It is forecasted that in 2037 the number of electric vehicles on the road will need an amount of electricity equal to the entire "green" electricity produced at that time. Therefore, "green" operation of electric vehicles entering the market after 2037 will be impossible. The sales of hydrogen-burning vehicles are poised to overtake the sales of electric vehicles in 2041, but a non-polluting overall operation for them is not guaranteed either at this time.

## 1. Introduction

Battery electric vehicles (BEVs) first came into existence in the late 19th century, but their numbers began growing significantly in the 2010s following government incentives to increase adoption in the United States and the European Union.[1][2] The main argument in their favor has always been their non-polluting operation.

More recently hydrogen vehicles have been introduced. Again the argument in their favor has been their non-polluting operation. A hydrogen vehicle is a vehicle that uses hydrogen fuel for motive power. Power is generated by converting the chemical energy of hydrogen to mechanical energy by reacting hydrogen with oxygen in a fuel cell to power electric vehicles (FCEVs.) In the last couple of years hydrogen vehicles powered by internal combustion engines (HICEVs) also made their appearance but their sales worldwide have so far achieved negligible market penetration. It has been suggested that sales of HICEVs will take off toward the end of the 2020s and will reach 400,000 shipments by 2040.[3] This eventuality seems rather hypothetical at this time.

Today the numbers of vehicles on the roads of the three types, namely, old internal combustion engine vehicles (ICEVs,) BEVs plus plug-in hybrid electric vehicles (PHEVs), and FCEVs, show populations smaller in steps of roughly one order of magnitude each. But BEVs have been gaining market share from ICEVs very rapidly.

In this work use will be made of the substitution model introduced by Cesare Marchetti and collaborators in order to analyze and forecast the transition between the three more important vehicle technologies mentioned earlier.[4] This model fits logistic curves on the data, which adds reliability to the forecasts considering that the logistic function describes the law of natural growth in competition. The results will then be confronted with a similar analysis and forecasts of the worldwide primary-energy consumption.

The fact that sales of non-polluting vehicles are growing much more rapidly than the rate at which production of "green" electricity is growing implies that electric vehicles will be using progressively more and more "black" electricity, i.e. electricity produced by burning fossil fuels. In fact there will be a time – estimated below as 2037 – when there will not be enough "green" electricity around for the number of electric vehicles in circulation. At that point in time non-polluting electric vehicles will see their primary *raison d'être* severely compromised.

## 2. The rush to non-polluting vehicles

The sales of BEVs + PHEVs have been growing rapidly in the last ten years, and they have been steadily taking market share away from the old ICEVs. In Figure 1 we see the evolution of the worldwide sales of the three categories of vehicles. The vertical scale is *logistic*, which transforms S-shaped logistic curves into straight lines by plotting the logarithm of the ratio $f/(1-f)$ where $f$ if the market share of the technology in question. The scale is highly non-linear with 0 at $-\infty$ and 100% at $+\infty$. The straight line labeled FCEV is not a fit to the two data points, but its slope is set – by analogy – equal to the magnitude of the slopes of the other straight lines in the drawing. The curved parts of the trajectories are calculated as 100% minus the market shares of all other technologies that are following straight-line trajectories. This methodology has been used extensively by Marchetti et al. at the International Institute for Applied Systems Analysis (IIASA.)[5][6]

We see in Figure 1 that the data points fall on straight lines, which means they are following S-shaped logistic trajectories. The data come from the International Energy Agency (IEA) and from Hydrogen Insight.[7][8] The two early technologies, ICEVs and (BEVs+PHEVs) will split the market equally around 2027. The electric vehicles (BEVs+PHEVs) will reach a maximum market share of 92% in 2034 when the other two – ICEVs and FCEVs – will split the remaining 8% of the market just about equally between them. After 2034 electric vehicles will begin losing market share to FCEVs and will end up splitting the market equally with them in 2041 when ICEVs will be reduced to a market share smaller than 1%. A scenario for a role played by HICEVs in the future is included (intermittent line.) Their eventual growth, if it becomes realized, will be at the expense of FCEVs.

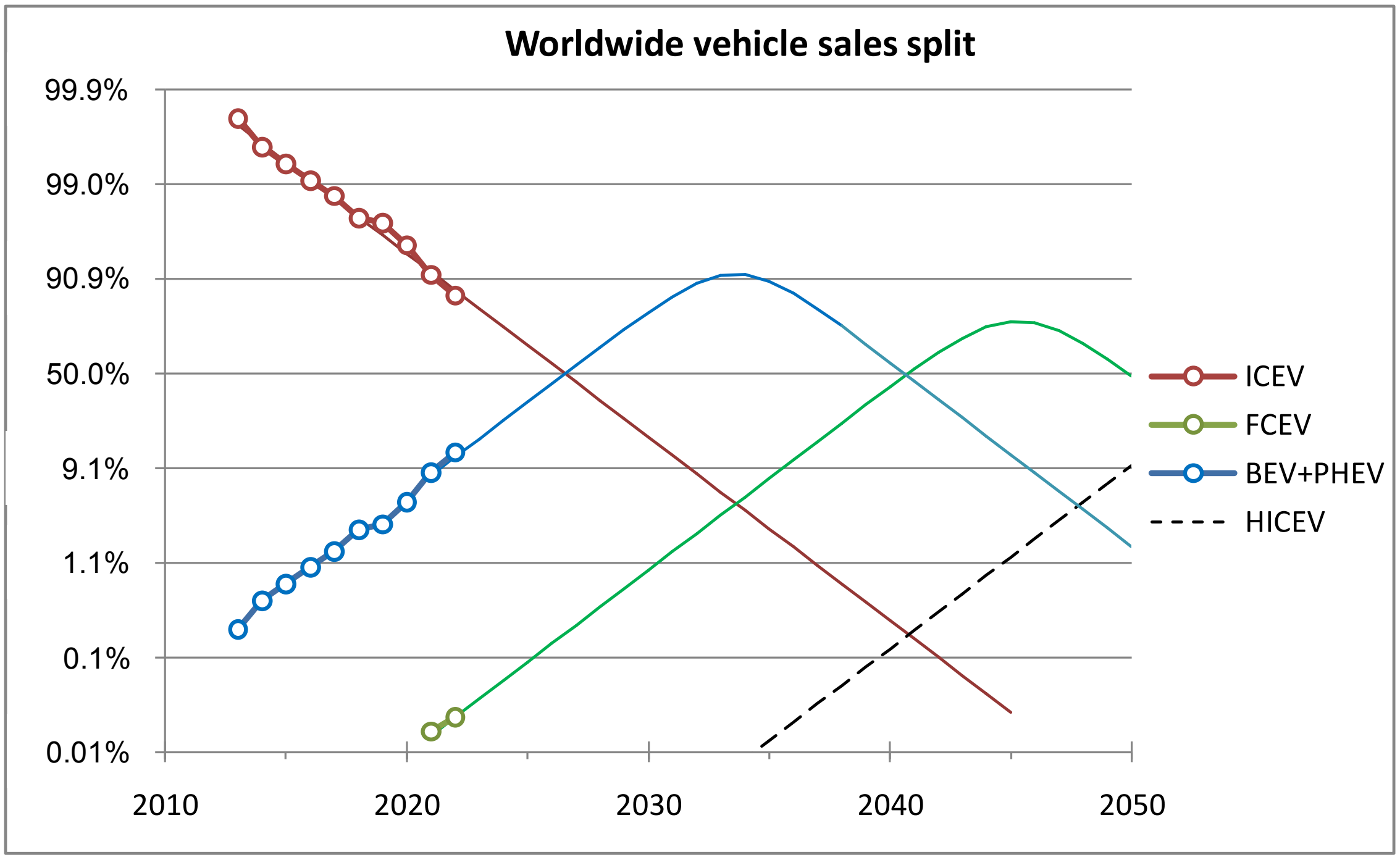

Figure 1. The vertical scale is non-linear (*logistic*), with 0 at -∞ and 100% at +∞, and shows market shares. Straight lines denote S-shaped natural-growth curves. The intermittent line constitutes a hypothetical scenario. Data source: International Energy Agency (IEA)

The market shares of Figure 1 can be translated to number of vehicles if we estimate the overall size of the market. In Figure 2 we see estimates of worldwide vehicle sales from three different sources (IEA, Statista, and ChatGPT.) These are rough estimates as indicated by the significant discrepancies between them. The thick pink line is an average, which is then extrapolated in a trivial way (intermittent pink line.) Despite the important uncertainties of the forecasts, Figures 3a, 3b, and 3c show revealing trends for the three types of vehicles mentioned in Figure 1.

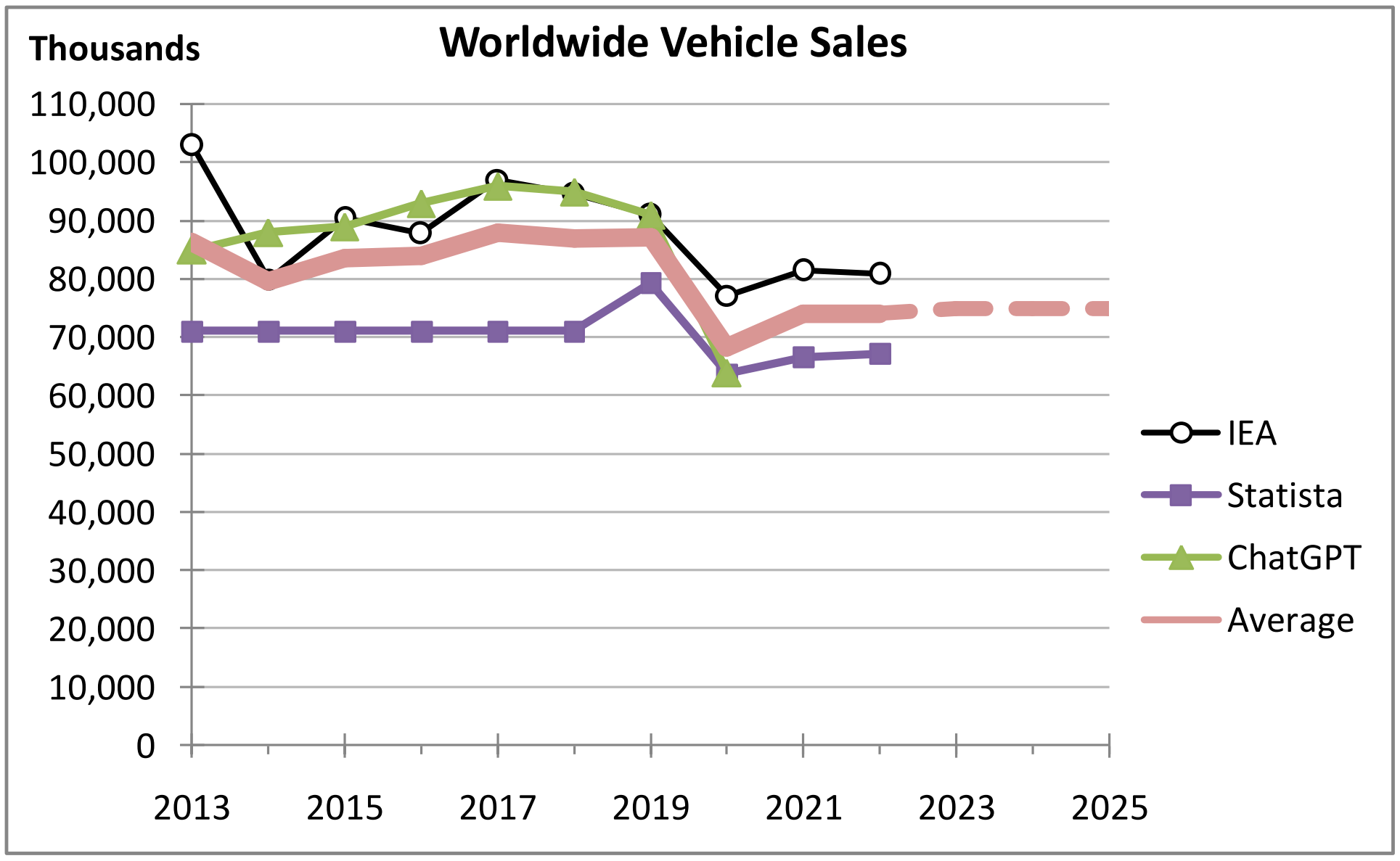

Figure 2. Estimates for worldwide vehicle sales. The average (pink line) is forecasted in a trivial way (intermittent pink line.) Data sources: sources (IEA, Statista, ChatGPT)

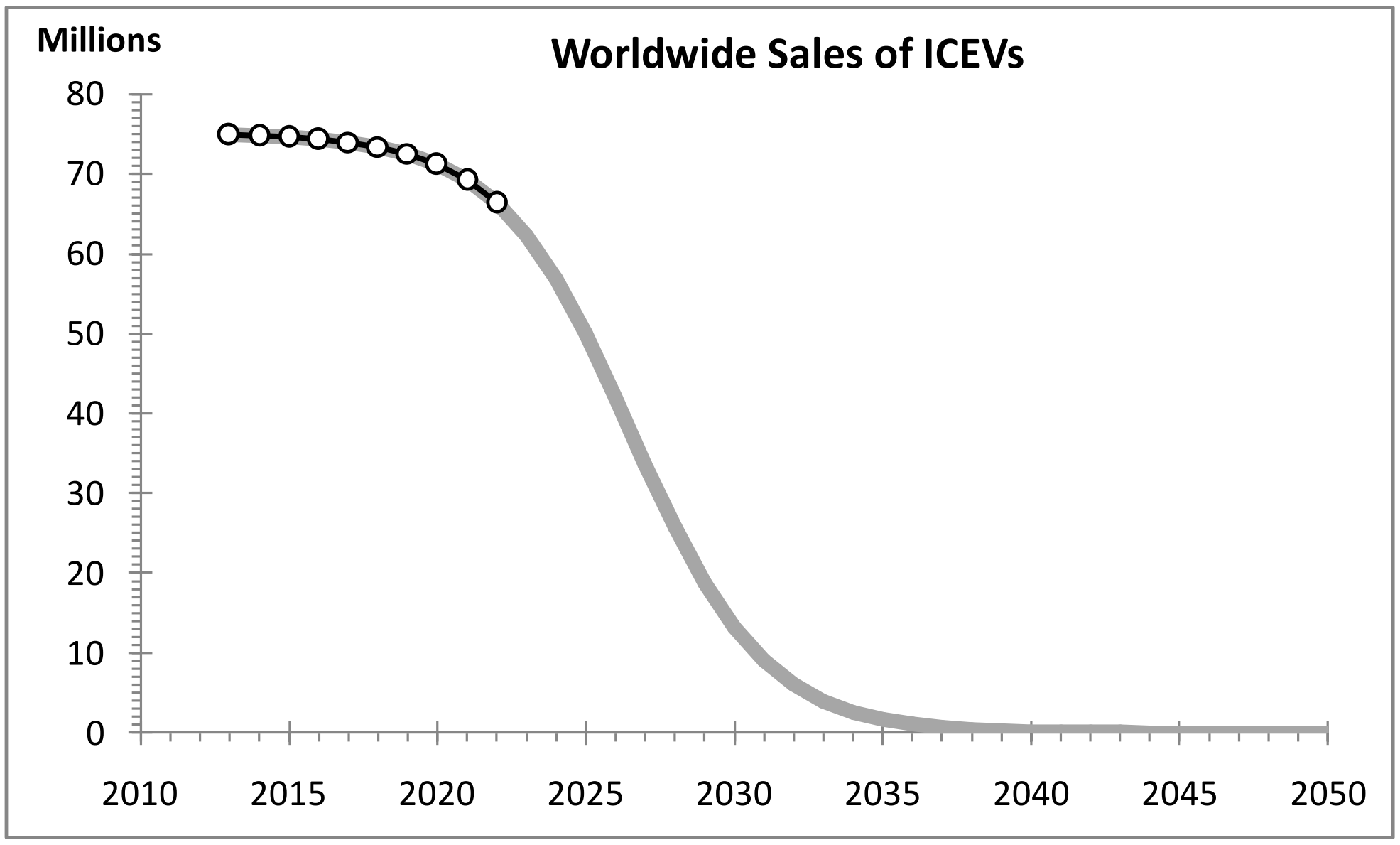

Figure 3a. The gray line is obtained by combining the forecasts in Figures 1 and 2.

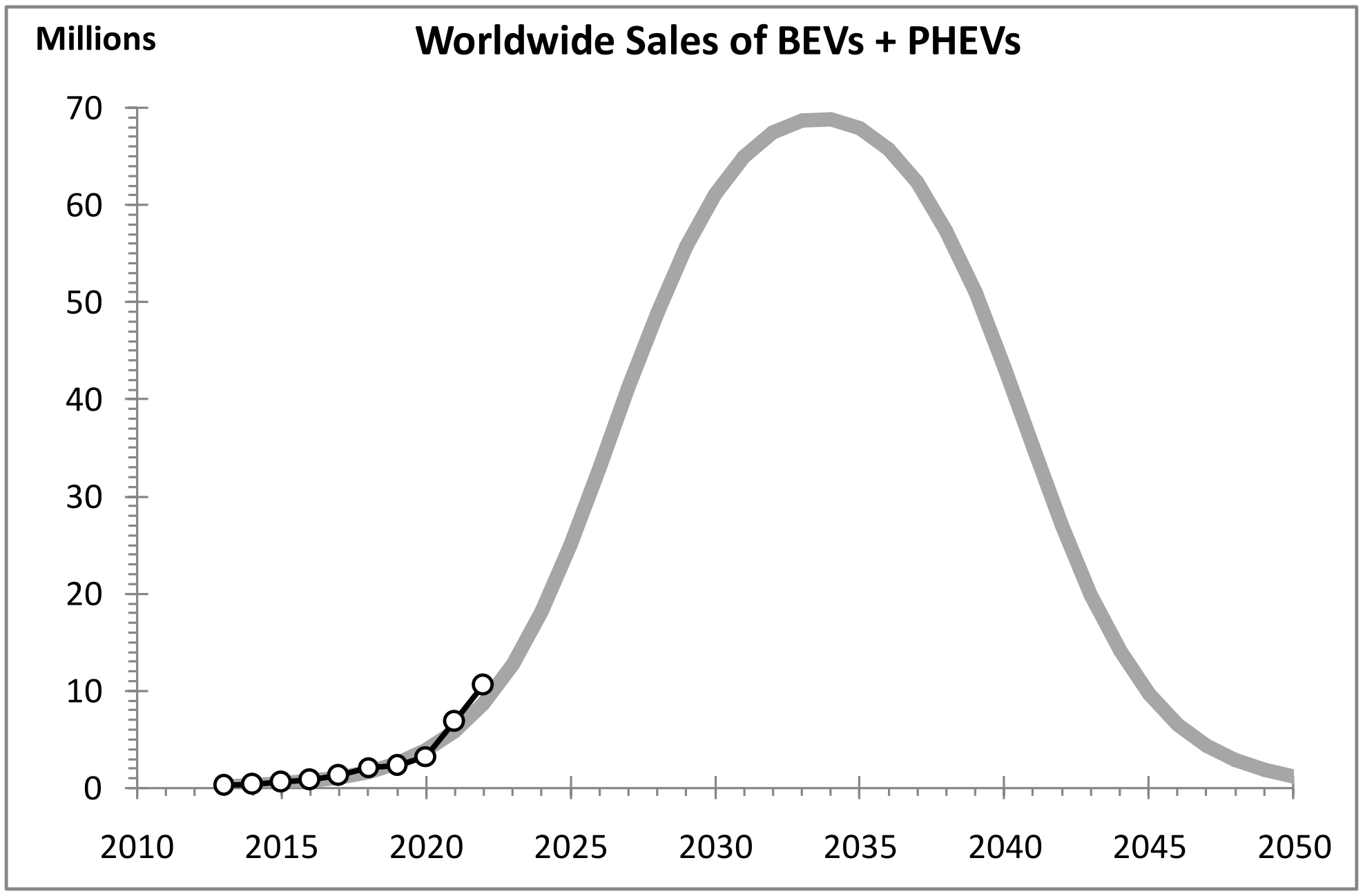

Figure 3b. The gray line is obtained by combining the forecasts in Figures 1 and 2.

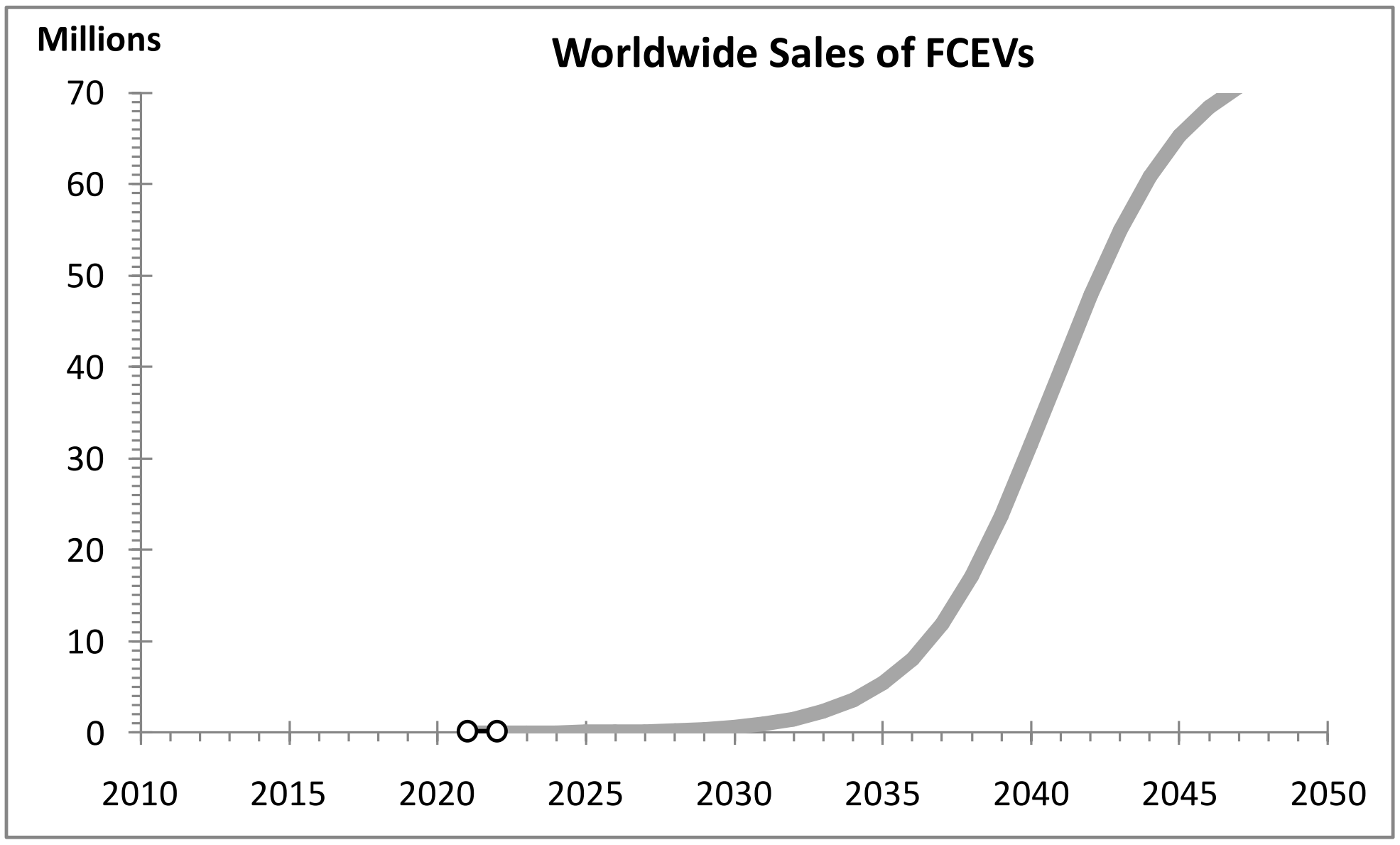

Figure 3c. The gray line is obtained by combining the forecasts in Figures 1 and 2.

## 3. Vehicles on the roads

The picture of the number of vehicles on the road is quite different from that of the number of vehicles being sold because people tend to keep their cars for a variable number of years. Combining data from IEA, Statista, and Wikipedia we obtain Figure 4 showing worldwide vehicle registrations. It is qualitatively similar to Figure 1, but there is a delay of 3 to 4 years, namely, the maximum market share (90%) for BEV+PHEV occurs in 2037 instead of 2034, and ICEVs decline to 50% and FCVEs reach 50% in 2031 and 2045 respectively instead of 2027 and 2041. The future eventuality of HICEVs is not considered here.

But in both pictures, i.e. sales and registrations, the onslaught of what is generally considered as non-polluting vehicles is unquestionable.

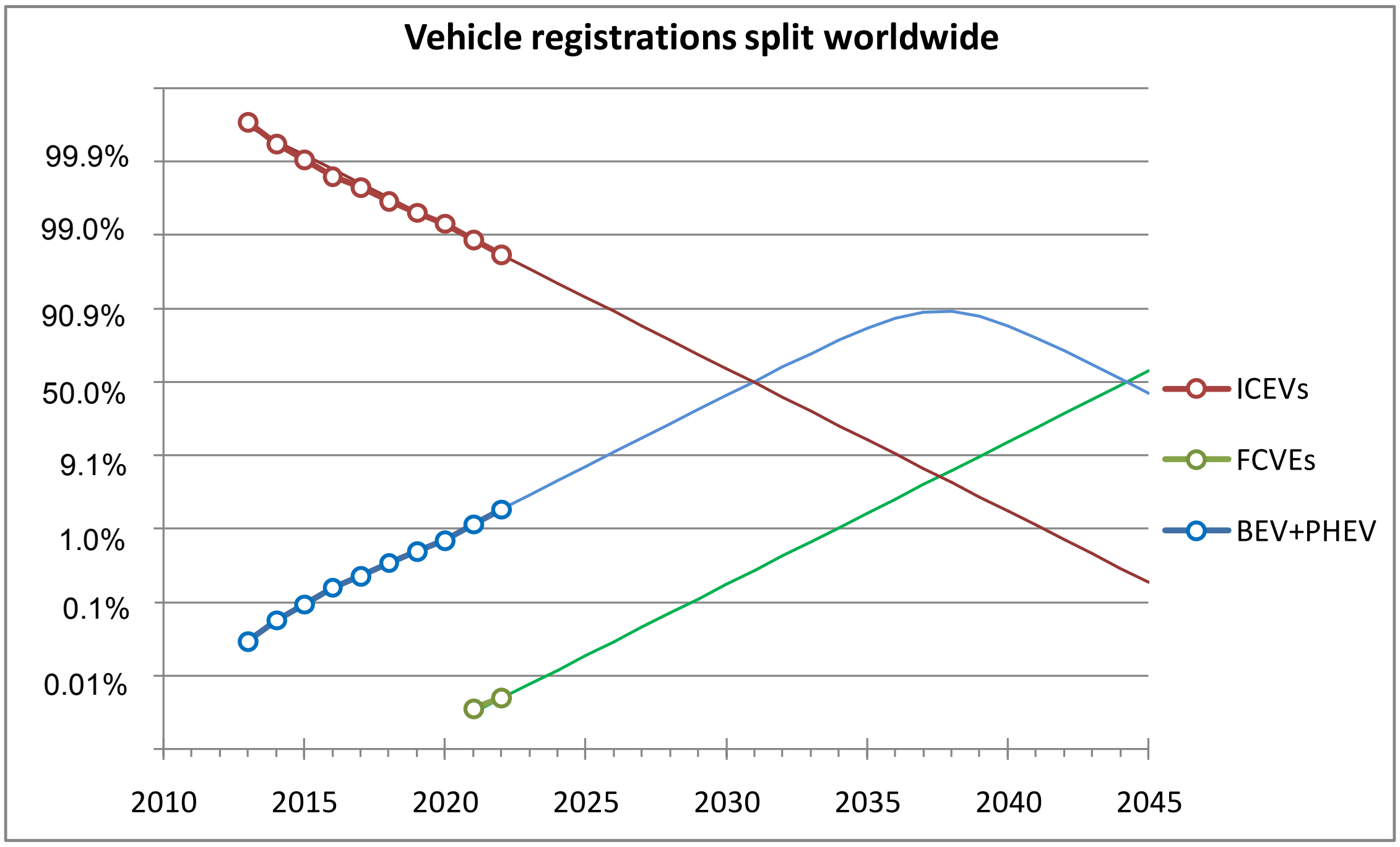

Figure 4. The vertical scale is non-linear (*logistic*), with 0 at -∞ and 100% at +∞, and shows market shares. Straight lines denote S-shaped natural-growth curves.
Data sources:IEA, Statista, and Wikipedia

### 3.1 The electric-vehicle micro niche BEVs+PHEVs

The electric-vehicle micro niche BEVs+PHEVs has been looked at more closely in order to identify the market dynamics between BEVs and PHEVs. Figure 5 shows that the two types of vehicles split this micro niche with flat comparable shares until 2016. But later, PHEVs began losing market share systematically. By 2035 their share is expected to be half of what it is today. This is compatible with the fact that the number and availability of charging stations is rapidly increasing worldwide thus undercutting the main advantage that PHEVs have over BEVs today, namely extended driving range (reduced range anxiety.)

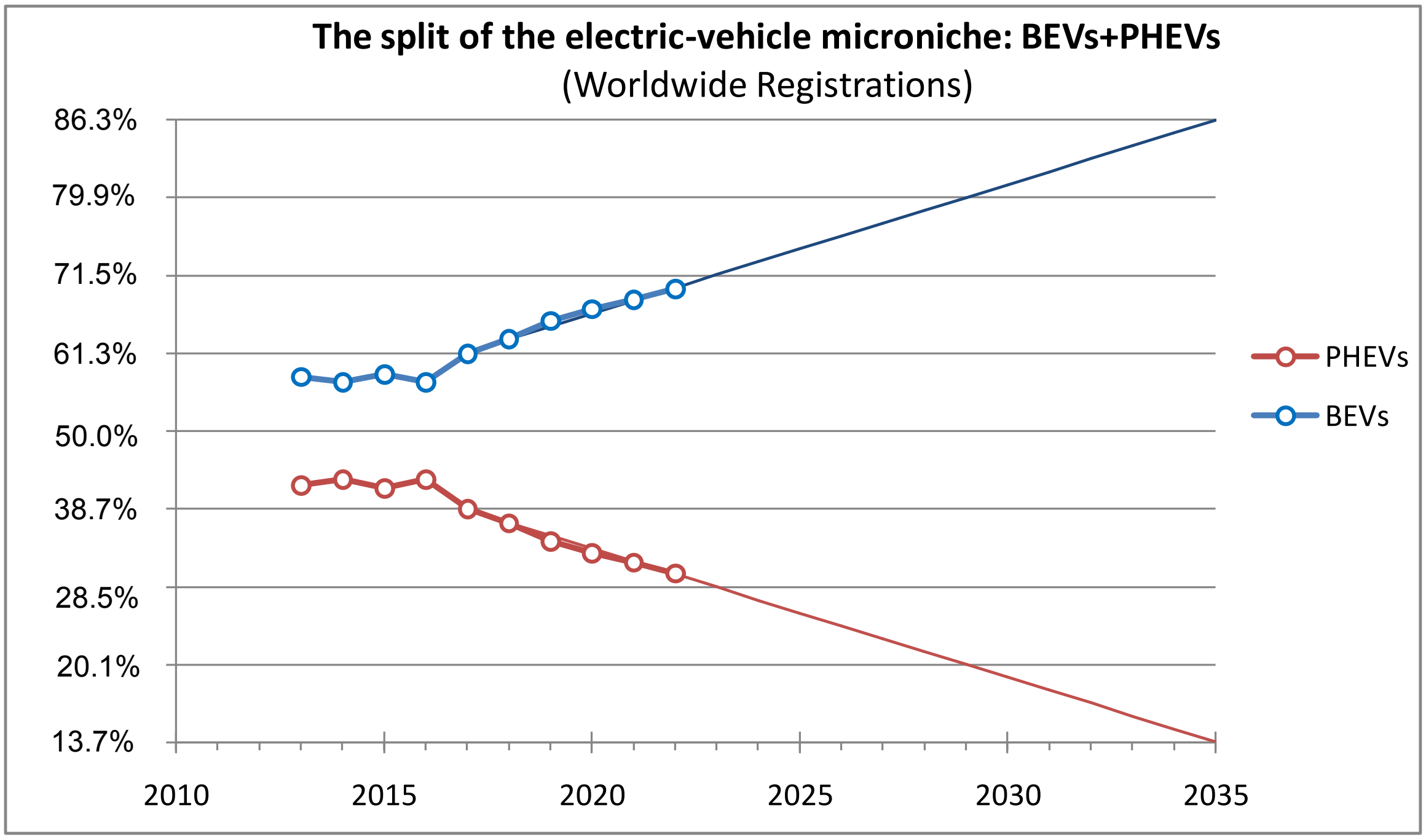

Figure 5. The vertical scale is non-linear (*logistic*), with 0 at -∞ and 100% at +∞, and shows shares of the microniche. Straight lines denote S-shaped natural-growth curves. Data source: IEA

## 4. "Green" electricity production

The world energy picture has been recently analyzed using the same methodology but focusing on the different types of renewable energies.[9][10] In Figure 6 we now focus on "green" electricity production coming from nuclear, renewables, and hydroelectric plants. We therefore consider a two-way split of all primary energies, namely, those that produce "green" electricity with no $CO_2$ emissions (nuclear, renewables, and hydroelectric), and those that produce "black" electricity burning fossil fuels (coal, oil, natural gas.)

The picture – Figure 6 – shows a clear substitution of non-polluting primary energies for polluting ones. Again, the nature of this substitution being a natural-growth process is evidenced by the fact that the percentage trajectories are amenable to descriptions by logistic functions. But the slopes are gentle, much gentler than the slopes in Figure 4. The mid-point of this substitution – when there will be equal amounts of polluting and non-polluting energies consumed – is estimated around 2102.

Another substitution process pertinent at this point is how electricity is produced worldwide, see Figure 7. Today 40% of the electricity produced is "green" while the rest is produced by burning fossil fuels. The projections estimate that half of the electricity produced will be "green" by 2036.

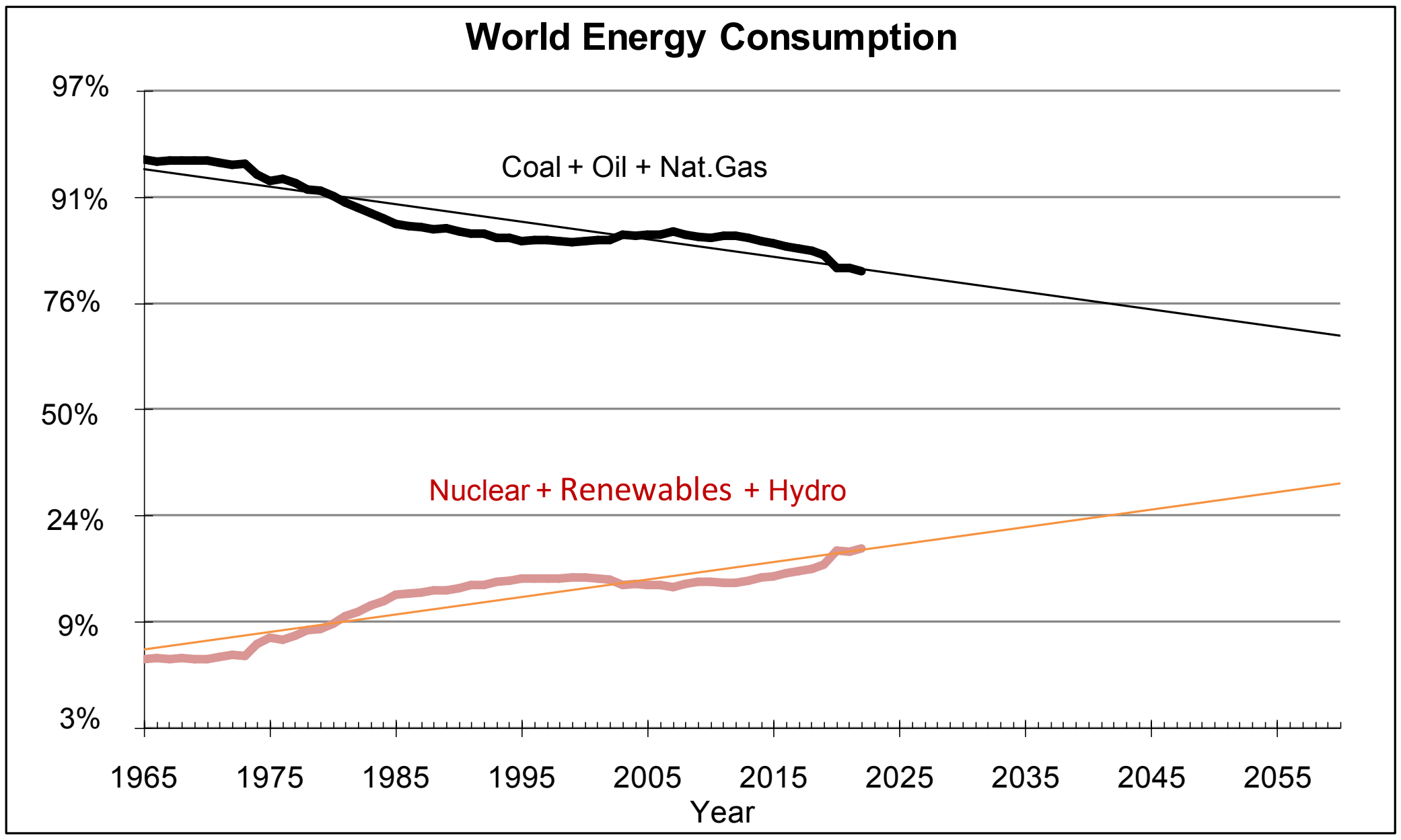

Figure 6. The vertical scale is non-linear (*logistic*), with 0 at -∞ and 100% at +∞, and shows market shares. Straight lines denote S-shaped natural-growth curves.
Data source: https://ourworldindata.org/energy

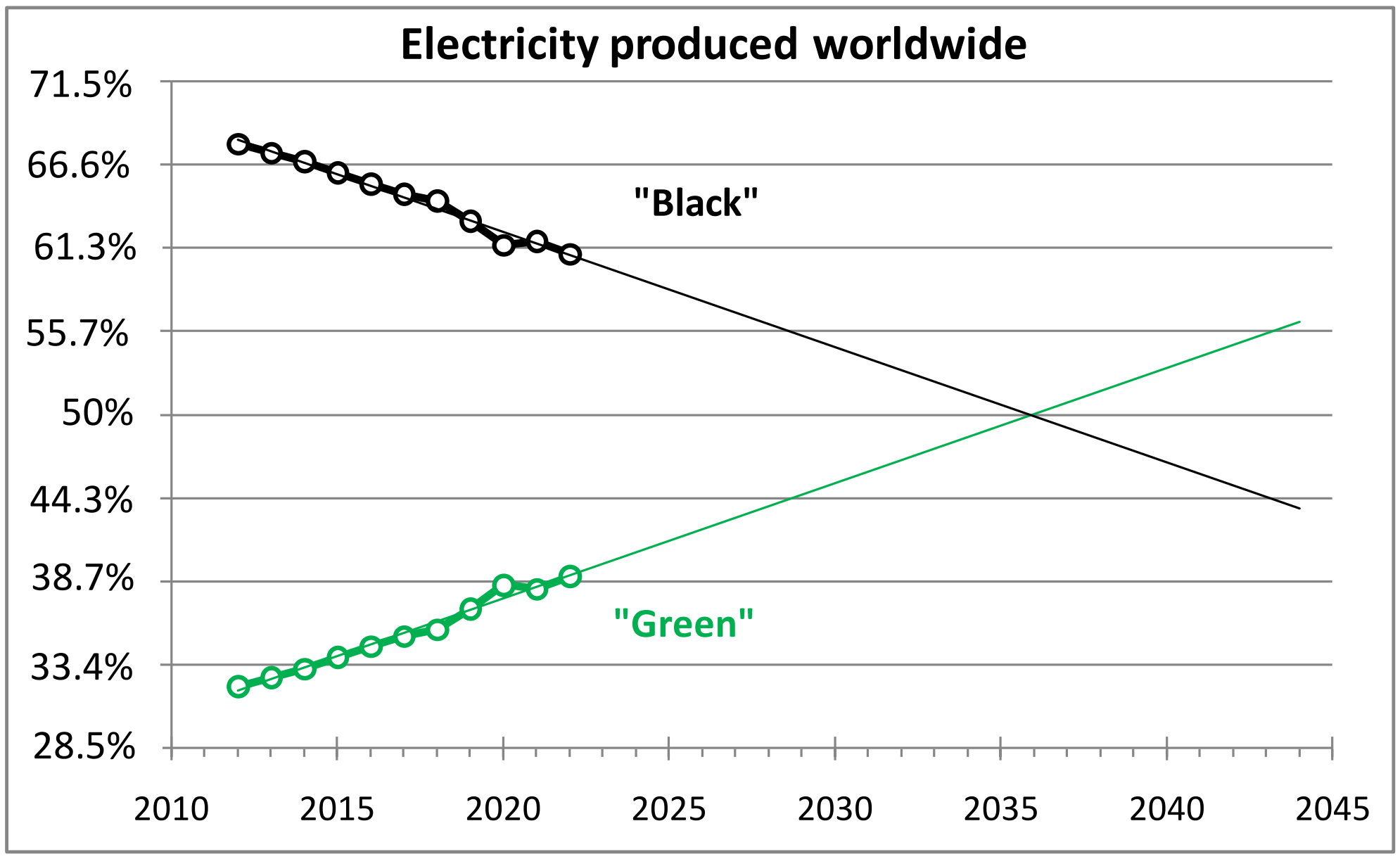

Figure 7. The vertical scale is non-linear (*logistic*), with 0 at -∞ and 100% at +∞, and shows market shares. Straight lines denote S-shaped natural-growth curves.
Data source: https://ourworldindata.org/energy

## 5. Discussion

The use of the logistic substitution model indicates that the various vehicle technologies are poised to substitute for one another successively: ICEVs to (BEVs+PHEVs) to FCEVs and possibly to HICEVs. But the hybrid technology (PHEVs) seems to be slowly falling out of favor with the public. The forecasts, obtained by fitting straight lines in *logistic* plots of the data – essentially fitting S-curves – enjoy enhanced validity because *logistic* patterns describe natural-growth processes. The forecasts for vehicle populations can now be contrasted to forecasts for "green" electricity production.

The Energy Information Administration has released data showing that the transportation of people and goods accounts for about 25% of all energy consumption in the world. Let us assume that this percentage will not change significantly in the future. From Figure 4 we see that by 2037, 90% of the vehicles on the road will be electric. This electric-vehicle fleet will need 25% x 90% = 22.5% of all energy consumed. But we see in Figure 6 that the "green" electricity produced in 2037 will represent only 22.7% of all energy produced. Therefore, by 2037 electric vehicles will demand practically all the "green" electricity produced. Consequently new electric vehicles entering the market from 2037 onward will be using electricity produced only by fossil burning, which challenges the main *raison d'être* of electric vehicles. Appropriately, we see in Figure 4 that the share of BEVs+PHEVs begins to decline around 2037 in favor of FCEVs, which do not require input of electricity for their operation.

As shown in Figure 7, today's electricity production is 60% "black" and is projected to diminish only to 50% by 2036. Therefore statistically speaking, electric vehicles are likely to burn "black" electricity more often than not and for a long time. In fact, things are even worse because of the habit – convenient to most people – to charge their vehicles during the night hours while they sleep. Electricity produced during the night is mostly "black."[13][14]

It has been argued that BEVs using electricity produced by burning fossil-fuel are still less polluting than ICEVs because of enhanced efficiencies in electric power plans due to economies of scale, and also because of advanced filtering of factory emissions. On the other hand, BEVs carry an additional pollution tag through the manufacturing of their massive batteries, which must also be taken into account. The detailed calculation of the equation becomes rather complicated, but studies show that BEVs may indeed be less polluting than ICEVs.[11][12]

It is expected that over time efficiencies of vehicle engines as well as electricity-producing factories will improve further, BEV batteries will become lighter, and fuel cells will become less dependent on precious metals. But all these technological improvements follow natural-growth processes that have been going on for a while and therefore their influence on the determination of the logistic trajectories in Figures 1, 4, 5, and 6 has already been taken into account.

A significant growth of HICEVs, as hypothesized in Figure 1, was not analyzed in this work even if they have some advantages over FCEVs. For example, they are less expensive to construct, and also, hydrogen has higher energy density than batteries used in FCEVs, which can result in longer driving ranges and shorter refueling times for HICEVs. But they don't enjoy a clear advantage in terms of a "greener" operation. They depend on hydrogen and the primary method for large-scale production of hydrogen is steam methane reforming (SMR), which produces a lot of $CO_2$ in the process. If ways are developed in the future to captured and dispose $CO_2$ effectively, or enough green energy becomes available to convert $CO_2$ into carbon-neutral hydrocarbons, then one can envisage HICEVs as a third entrant in Figure 4 around the mid 2030s growing at the expense of FCEVs causing their trajectory to go over a maximum like the one shown earlier for BEVs+PHEVs.[15]

In a nutshell, the bottleneck today is not the vehicle technology but rather the availability of “green” energy. In the fight against climate change society’s efforts have been concentrated too much on the development of vehicle technologies and not enough on the production of “green” energy. As this gap increases, embarrassing situations can arise, such as the production of fancy cars that have little to show in terms of improving climate change.

**Author statement**

I would like to thank Athanasios G. Konstandopoulos for fruitful discussions.